\documentclass[aps,prd,twocolumn, showpacs, nofootinbib]{revtex4-1}
\usepackage{bm}
\usepackage{amsmath}
\usepackage{amsfonts}          
\usepackage{latexsym}
\usepackage{amssymb}
\usepackage{amsmath}
\usepackage{dsfont}
\usepackage{url}
\usepackage{graphicx}
\usepackage[applemac]{inputenc}  
\usepackage[english]{babel}
\usepackage{color}

\numberwithin{equation}{section}

\newcommand{\be}{\begin{equation}}
\newcommand{\ee}{\end{equation}}

\newcommand{\p}{\bm{p}}
\newcommand{\cz}{\bm{\chi}_0}
\newcommand{\cu}{\bm{\chi}_1}
\newcommand{\cd}{\bm{\chi}_2}
\newcommand{\np}{(\bm{n}\cdot \bm{p})}
\newcommand{\ncz}{(\bm{n}\cdot \bm{\chi}_0)}
\newcommand{\ncu}{(\bm{n}\cdot \bm{\chi}_1)}
\newcommand{\ncd}{(\bm{n}\cdot \bm{\chi}_2)}
\newcommand{\pcz}{(\bm{p}\cdot \bm{\chi}_0)}
\newcommand{\pcu}{(\bm{p}\cdot \bm{\chi}_1)}
\newcommand{\pcd}{(\bm{p}\cdot \bm{\chi}_2)}

\newcommand{\cesq}{\left(\bm{\chi}^2 \right)_\text{eff}}
\newcommand{\ncesq}{(\bm{n} \cdot \bm{\chi})_\text{eff}^2}
\newcommand{\pcesq}{(\bm{p} \cdot \bm{\chi})_\text{eff}^2}
\newcommand{\npcesq}{\left((\bm{n} \cdot\bm{\chi})(\bm{p} \cdot \bm{\chi}) \right)_\text{eff}}
\newcommand{\enpcesq}{\left(\left( \bm{n} \times \bm{p} \right) \cdot \bm{\chi}\right)^2_\text{eff}}

\newcommand{\mud}{\frac{m_1}{m_2}}
\newcommand{\mdu}{\frac{m_2}{m_1}}

\newcommand{\hhef}{\hat{H}_\text{eff}}
\newcommand{\gef}{g_\text{eff}}

\begin{document}

\title{Effective-one-body Hamiltonian with next-to-leading order spin-spin coupling} 
\author{Simone Balmelli}
\email{balmelli@physik.uzh.ch}
\author{Philippe Jetzer}

\affiliation{Physik-Institut, Universit\"at Z\"urich, Winterthurerstrasse 190, 8057 Z\"urich, Switzerland}
\begin{abstract}
We propose a way of including the next-to-leading (NLO) order spin-spin coupling into an effective-one-body (EOB) Hamiltonian.
This work extends [S. Balmelli and P. Jetzer, Phys. Rev. D \textbf{87}, 124036 (2013)], which is restricted to the case of equatorial orbits and aligned spins, to general orbits with arbitrary spin orientations. 
This is done applying appropriate canonical phase-space transformations to the NLO spin-spin Hamiltonian in Arnowitt-Deser-Misner (ADM) coordinates, and systematically adding  ``effective'' quantities at NLO to all spin-squared terms appearing in the EOB Hamiltonian.
As required by consistency, the introduced quantities reduce to zero in the test-mass limit.
We expose the result both in a general gauge and in a gauge-fixed form.
The last is chosen such as to minimize the number of new coefficients that have to be inserted into the effective spin squared.
As a result, the 25 parameters that describe the ADM NLO spin-spin dynamics get condensed into only 12 EOB terms.

\end{abstract}


\maketitle

\section{Introduction}
Thanks to the LIGO/Virgo network of second-generation ground based interferometers, a first direct detection of gravitational waves (GW) is expected to occur in few years \cite{aas:11}. 
Furthermore, in the next decades, space-born GW detectors \cite{vita:14} (such as the planned eLISA) will open an entire new window to astrophysics, and allow tests of General Relativity at an unprecedented level \cite{amar:13}. 
Both types of detectors rely on coalescing (and possible spinning) black hole binaries as a primary GW source.

In order to extract the GW signal from the noise, very precise waveform templates need to be constructed. 
Currently, the most accurate description of coalescing black holes is provided by numerical relativity (NR) (see e.g. \cite{lov:12, buch:12, mrou:13, hind:14} for some recent advances).
However, the full parameter space is too large (especially in the case of nonzero spins) for being densely covered by NR simulations .
This is the main reason why semi-analytical methods, by far less computationally expensive, can turn out to be very useful.
At the present time, the effective-one-body (EOB) approach (we refer to \cite{dam:12_EOB} for a general review) is the only semi-analytical method that has been able to accurately describe the complete waveform of a coalescing process. 
After years of constant development \cite{buon:99, buon:00, dam:00, dam:02, buon:03, dam:07, dam:07_2, buon:07, buon:07_2, dam:08_4, dam:08_5, dam:08_3, dam:09, buon:09, dam:09_2, dam:12_EvsJ} the EOB has now reached an excellent agreement with NR waveforms in the case of nonspinning binaries \cite{dam:13, hind:14}.
A lot of effort has also been put into the modeling of spins \cite{dam:01, dam:08, bar:09, bar:10, pan:10, bar:11, pan:11, pan:11_2, nag:11, tar:12, bal:13, bal:13:err, tar:14, pan:14, dam:14}, leading to a good overlap with NR waveforms in the case of nonprecessing (aligned or anti-aligned) spins \cite{hind:14, tar:14, dam:14}.
By contrast, the description of precessing spins still needs some work before reaching a comparable performance \cite{pan:14}, and is currently one of the most urgent tasks. 
In view of this, it may be crucial to incorporate more analytical information from the spin-orbit and spin-spin Hamiltonian computed within the post-newtonian (PN) theory. 

In Ref.~\cite{bal:13}, a possible way of including the next-to-leading (NLO) spin-spin coupling \cite{stein:08s1s2, stein:08s^2} (see also \cite{por:08, por:10, por:08_2, por:10_2, levi:10, levi:14}) into the spinning EOB model of Refs.~\cite{dam:01,dam:08,nag:11} has been exposed for the special case of equatorial orbits and nonprecessing spins. 
A general inclusion of NLO spin-spin effects for fully precessing orbits would be a necessary step for extending the reliabilty of EOB waveforms to a significantly larger portion of the parameter space. 
The present paper aims at filling this gap, providing a possible implementation of the missing terms.

Recently, an improved (and calibrated)  EOB model for spinning binaries has been proposed \cite{dam:14}, where the NLO spin-spin coupling is only incorporated for the case of circular orbits. 
The present paper could be a first step for developing, at a next stage, a more general version of that improved model.

The paper has the following structure: in Sec.~(\ref{sec:summary}) we summarize the main concepts and the formalism of Ref.~\cite{bal:13}. 
In Sec.~(\ref{sec:inclusion}), which is the central part of the paper, we propose a way of including the NLO spin-spin terms and show the explicit result, both in a general form (with a given number of free gauge parameters) and in a gauge-fixed formulation. 
Finally, Sec.~(\ref{sec:discussion}) discusses a second, slightly different way of incorporating the wished spin-spin terms. 
Both approaches are compared plotting the (gauge invariant) angular frequency and binding energy at the last stable orbit (LSO).
The plot also shows the prediction of the calibrated models of Refs.~\cite{tar:12, tar:14}.

Throughout the paper, we use geometric units with $G \equiv c \equiv 1$.
When writing formulae in a PN expanded form, however, we will reintroduce the usage of $c$, prepending a factor $\left(1/c\right)^{2n}$ with the mere purpose of labeling the PN order $n$.

\section{Summary of the previous work}
\label{sec:summary}

In this section, we outline the method followed in Ref.~\cite{bal:13}, which forms the basis of the current paper.
We work in EOB coordinates, with $R$ being the radial coordinate, $\bm{n}$ the unit radial vector and $\bm{P}$ the momentum vector.
We will often use the rescaled variables $r \equiv R/M$ and $\bm{p} \equiv \bm{P}/\mu$.  
Here, $M \equiv m_1 + m_1$ is the central EOB mass ($m_1$ and $m_2$ being the individual masses of the two black holes), and $\mu \equiv m_1\,m_2/M$ is the reduced mass.
Moreover, we denote by $ \nu \equiv \mu/M$ the symmetric mass ratio.

The starting point is the EOB model of Ref.~\cite{dam:01} (which includes both spin-spin and spin-orbit coupling at leading order (LO)), together with its extensions to the NLO \cite{dam:08} and to the next-to-next-to leading order (NNLO) \cite{nag:11} spin-orbit coupling. 
As already mentioned, in Ref.~\cite{bal:13} the NLO spin-spin Hamiltonian in ADM coordinates has been reformulated and inserted into this EOB model for the special case of two black holes whose spins are aligned (or anti-aligned) with the orbital angular momentum.
In particular, it has been shown that it is sufficient to replace the spin parameter $a_0$ of the effective metric (see for instance Eqs.~(4.7) and (4.8) of Ref.~\cite{bal:13}), whenever it appears as a second power, by a new, effective \emph{squared} spin parameter.
Using the dimensionless notation $\chi_0 =  a_0 / M$, this prescription takes the form 

\begin{equation}
	\label{eq:al_prescr}
	\chi_0^2 \to (\chi^2)_\text{eff} = \chi_0^2 + \Delta \chi_\text{eff}^2.
\end{equation}
We recall that $\chi_0$ is a combination of the dimensionless spin parameters  $\chi_1$ and $\chi_2$ ($ \chi_a = S_a /m_a^2 $) of the two bodies:

\begin{equation}
\chi_0 = \frac{m_1}{M} \chi_1 + \frac{m_2}{M} \chi_2.
\end{equation}
The additional term $\Delta \chi_\text{eff}^2$ is of fractional 1PN order with respect to $\chi_0^2$ and carries the information for reproducing the correct NLO spin-spin coupling.
It reads as

\begin{align}
	\label{eq:dchieff}
	\Delta \chi_\text{eff}^2 = & \frac{1}{c^2}\bigg[\left( a_{11}\bm{p}^2  + \frac{c_{11}}{r}\right)\chi_1^2\nonumber\\
				& \quad +\left( a_{22}\bm{p}^2  + \frac{c_{22}}{r}\right)\chi_2^2\nonumber\\
				& \quad +\left( a_{12}\bm{p}^2  + \frac{c_{12}}{r}\right)\chi_1 \chi_2\bigg].
\end{align}

The calculation of the coefficients $a_{ab}$ and $c_{ab}$ is the main result of Ref.~\cite{bal:13}, given by Eq.~(5.12) there.
All of them vanish in the test mass limit $\nu \to 0$, consistently with the requirement that the EOB metric must reduce to the Kerr one.

PN results in ADM coordinates can be included into an EOB model after suitable canonical transformations.
We denote here by $G_\text{o}^\text{PN}$ and $G_\text{ss}^\text{PN}$ the generating functions of the corresponding purely orbital and spin-spin transformation, respectively.
The procedure for transforming the NLO spin-spin Hamiltonian from ADM into EOB coordinates is made of three steps:\footnote{Here, we treat the PN expansion under the assumption of rapidly rotating black holes ($S_a = \frac{m_a^2}{c} \chi_a$, with $| \chi_a | \lesssim 1$), which assigns a well-defined PN order to the spin-dependent terms.
As a consequence, throughout this paper, $G_\text{ss}$ and $H_\text{ss}$ are of 2PN order when labeled with ``LO'', of 3PN order when labeled with ``NLO'', and so on.}

\begin{itemize}
\item[1)] A purely orbital transformation
	\begin{equation}
	\label{eq:1sttransf}
		H_\text{ss}^{\text{NLO} \prime} = H_\text{ss}^\text{NLO(ADM)} + \big\{G_\text{o}^\text{1PN}, H_\text{ss}^\text{LO(ADM)} \big\}.	
	\end{equation}
\item[2)] A LO spin-spin transformation
	\begin{equation}
	\label{eq:2dtransf}
		H_\text{ss}^{\text{NLO} \prime \prime } = H_\text{ss}^{\text{NLO} \prime} + \big\{G_\text{ss}^\text{LO}, H_\text{o}^{\text{1PN}\prime} \big\},
	\end{equation}
	where
	\begin{equation}
	\label{eq:2transf.2}
		H_\text{o}^{\text{1PN}\prime} = H_\text{o}^\text{1PN(ADM)} + \big\{G_\text{o}^\text{1PN}, H_\text{o}^\text{N(ADM)} \big\}.
	\end{equation}
\item[3)] A NLO spin-spin transformation
	\begin{equation}
	\label{eq:3dtransf}
		H_\text{ss}^{\text{NLO} \prime \prime \prime} = H_\text{ss}^{\text{NLO} \prime \prime} + \big\{ G_\text{ss}^\text{NLO}, H_\text{o}^\text{N(ADM)} \big\}.
	\end{equation}
\end{itemize} 

Notice that, for consistency, the transformations must be performed in a well-defined order.
As indicated above, we make \emph{first} use of the orbital transformation $G_\text{o}$, and \emph{then} of the spin-spin transformation $G_\text{ss}$
\footnote{Since the set of canonical transformations carries a group structure, the successive evaluation of $G_\text{o}(\bf{r},\bf{p}')$ and of $G_\text{ss}(\bf{r}',\bf{p}'')$ is a canonical transformation itself (with generating function $G_\text{ss,o}(\bf{r},\bf{p}'') = G_\text{o}(\bf{r},\bf{p}') + G_\text{ss}(\bf{r}',\bf{p}'') - \bf{r}'\cdot \bf{p}'$).
Despite taking a unique generating function would avoid the necessity of fixing an evaluation order prescription, we prefer here to use two separated transformations, so as to mantain the continuity with respect to Ref.~\cite{bal:13}.
As a second reason, the transformation of the Hamiltonian is more easily calculated here than for a single generating function, since in that last case effects quadratic in the transformation must be considered (see e.g. Eq.~(6.9) of Ref.~\cite{buon:99}).}. 
Notice that, in Ref.~\cite{nag:11}, the spin-orbit generating function $G_\text{so}$ is also applied after $G_\text{o}$.
By contrast, an evaluation order prescription between spin-orbit and spin-spin transformation would first be necessary when taking into account contributions that are cubic in the spins.
 
The final Hamiltonian must be equal to the corresponding term arising from the PN expansion of the EOB Hamiltonian.
Since, for the moment, this is only true in the spin-aligned case, we are just allowed to write:
\begin{equation}
	H_\text{ss,al}^{\text{NLO} \prime \prime \prime}= H_\text{ss,al}^\text{NLO(EOB)}.
\end{equation}

The NLO spin-spin transformation $\hat{G}_\text{ss,al}^\text{NLO} \equiv G_\text{ss,al}^\text{NLO}/ \mu$ is given by

		\begin{alignat}{2}
			\label{eq:GssNLOal}
 			 \hat{G}_{\text{ss,al}}^{\text{NLO}}=  \frac{\np}{c^6\, r}\bigg\{  & \bigg [ &&  \alpha_{11} \, \bm{p}^2 
					+\left(\gamma_{11}-\frac{1}{2}\right) \frac{1}{r} \bigg]  \bm{\chi}_1^2\nonumber\\
				+ &  \bigg [&& \alpha_{22} \, \bm{p}^2 
					 +\left(\gamma_{22}-\frac{1}{2}\right) \frac{1}{r}\bigg]  \bm{\chi}_2^2\\
			 	+ &  \bigg [ && \alpha_{12} \, \bm{p}^2 
					 +\gamma_{12} \frac{1}{r} \bigg]  (\bm{\chi}_1\cdot \bm{\chi}_2)\nonumber\bigg\}.
		\end{alignat}
Notice that, in view of the generalization to precessing orbits, we have written the individual dimensionless spins $\bm{\chi}_a = \bm{S}_a/m_a^2$ as vectors.
The coefficients $\alpha_{ab}$ and $\gamma_{ab}$ can be found in Eq.~(5.13) of Ref.~\cite{bal:13}.
It is also provided an expression for $\hat{G}_\text{ss}^\text{NLO}$ in the test mass limit (assuming $m_1 > m_2$):

\begin{equation}
	\label{eq:GssNLOKerr}
	\begin{split}
 		\lim_{\nu \to 0} \hat{G}_{\text{ss}}^\text{NLO} = \frac{1}{c^6 r^2}\bigg [ &-\frac{1}{2} \big (\bm{\chi_1}^2 +(\bm{n}\cdot \bm{\chi_1})^2 \big )(\bm{n} \cdot \bm{p}) \\
			& +  (\bm{p}\cdot \bm{\chi_1} )(\bm{n}\cdot \bm{\chi_1}) \bigg ].
	\end{split}
\end{equation}

The purpose of this paper is that of generalizing the prescription (\ref{eq:al_prescr}) to the case of general orbits, i.e., when the scalar products $\left( \bm{n} \cdot \bm{\chi}_a \right)$ and $\left( \bm{p} \cdot \bm{\chi}_a \right)$ ($a=1,2$) cannot be set to zero.
\section{Including NLO spin-spin terms into the EOB for general orbits}
\label{sec:inclusion}

\subsection{The prescription}

In Ref~\cite{bal:13}, the EOB metric is written in Boyer-Lindquist-like coordinates.
When spin precessions must be taken into account, however, it is necessary to switch to another system of coordinates. 
Following Ref.~\cite{dam:01} (and reformulating the angular variable $\theta$ according to $a_0 \cos( \theta) \equiv  ( \bm{n} \cdot \bm{a}_0) $), we can write the effective metric in Cartesian-like coordinates,
\begin{alignat}{2}
	\gef^{00} & = &&\frac{1}{\rho^2}\left(  \bm{a}_0^2 - \left(\bm{n} \cdot \bm{a}_0 \right)^2 - \frac{\left(R^2 +a_0^2\right)^2}{\Delta_t}\right)\\
	\gef^{0i} & = &&\frac{R}{\rho^2} \left(1 -\frac{R^2 + a_0^2}{\Delta_t} \right)\left(\bm{a_0}\times \bm{n}\right)^i \\
	\gef^{ij} & =&&\frac{1}{\rho^2} \bigg( \Delta_R n^i n^j + R^2\left(\delta^{ij} - n^i n^j\right) \\
		& && - R^2 \frac{ \left( \bm{a}_0 \times \bm{n} \right)^i \left( \bm{a}_0 \times \bm{n} \right)^j}{\Delta_t} \bigg),
\end{alignat}

where $\rho^2 = R^2 + \left( \bm{n} \cdot \bm{a}_0 \right)^2$.
$\Delta_t$ and $\Delta_R$ can be found e.g. in Eq.~(4.9) of Ref.~\cite{bal:13} (notice that they both depend on the spin through a term $\sim a_0^2$).
The effective Hamiltonian (that we denote here as ``old'', in order to avoid confusion with the modified version that is presented in this paper) takes the form
	\begin{equation}
		\label{eq:heffold}
		H_{\text{eff}}^\text{old} =  \Delta H_\text{so} + \beta^{i} P_{i} +   \alpha \, \sqrt{\mu^2 + \gamma^{ij}\,P_i\,P_j +Q_4(P_i)} ,  
	\end{equation}	  
with  a quartic-in-momenta term $Q_4(P_i)$ \cite{dam:00, dam:01} and with
	\begin{subequations}	
		\begin{alignat}{1}
			\alpha &= \frac{1}{\sqrt{- g_{\text{eff}}^{00}}} \\
			\beta^i &= \frac{g_{\text{eff}}^{0i}}{g_{\text{eff}}^{00}}\\
			\gamma^{ij} &= g_{\text{eff}}^{ij} + \frac{\beta^i \beta^j}{\alpha^2}.
		\end{alignat}	
	\end{subequations}
$\Delta H_\text{so}$ has been introduced to describe higher-order spin-orbit couplings.
It is defined in terms of the gyro-gravitomagnetic ratios $g_S^\text{eff}$ and $g_{S^*}^\text{eff}$, see Eqs.~(4.15) and (4.16) of Ref.~\cite{dam:08}.

With the reduced quantities $\cz = \bm{a}_0/M = (m_1 \bm{\chi}_1 + m_2 \bm{\chi}_2)/M$, $\hat{H} = H/\mu$, $\hat{\Delta}_r = \Delta_R/M$ and $\hat{\Delta}_t = \Delta_t/M$, we can write, more explicitly:

\begin{widetext}
\begin{subequations}
\label{eq:heffexpl}
\begin{align}
\label{eq:DHbetap}
	\Delta \hat{H}_\text{so} + \beta^i p_i  	&=   	\frac{r \nu}{2\tilde{r}^4}(r^2 + \cz^2 - \hat{\Delta}_t) \left(\left( \mud g_S^\text{eff} + g_{S^*}^\text{eff}\right)(\bm{n} \times \p) \cdot \cu+\left( \mdu g_S^\text{eff} + g_{S^*}^\text{eff}\right)(\bm{n} \times \p) \cdot \cd \right)\\
\label{eq:alpha}
	\alpha  				&=	 \left(\frac{\hat{\Delta}_t\left( r^2 + \ncz^2\right)}{\tilde{r}^4} \right)^{1/2} \\
\label{eq:gpp}
	\gamma^{ij}\,p_ip_j 		&=	 \frac{r^2}{r^2 + \ncz^2}\Bigg[\p^2 + \left( \frac{\hat{\Delta}_r}{r^2}-1\right) \np^2 - \frac{1}{\tilde{r}^4}\left( 2r^2 - \hat{\Delta}_t + \cz^2 + \ncz^2 \right) \left(\left( \bm{n} \times \bm{p} \right) \cdot \cz\right)^2 \Bigg],
\end{align}
\end{subequations}
\end{widetext}
where
\begin{equation}
\tilde{r}^4 = \left( r^2 + \cz^2 \right)^2- \hat{\Delta}_t \left(\cz^2 - \ncz^2\right).
\end{equation}
The spin-squared term $\left(\left( \bm{n} \times \bm{p} \right)\cdot \cz\right)^2$, generated by the contraction of $\gamma^{ij} p_i p_j$, can be expressed through the simple scalars $\p^2$, $\np$, $\cz^2$, $\ncz$ and $\pcz$ according to 
\begin{alignat}{2}
\label{eq:mixed_pr_id}
	 \left( \left( \bm{n} \times \bm{p} \right)\cz\right)^2 &= && \left( \p^2 - \np^2 \right)\left( \cz^2 - \ncz^2 \right) \nonumber \\
		& && - \left( \pcz - \np \ncz \right)^2. 
\end{alignat}

Let us now do some considerations:

\begin{itemize}
\item[i)] The prescription (\ref{eq:al_prescr}) acts selectively - leaving all non-squared spins untouched - and cannot be truly considered as a redefinition of the effective spin of the EOB  metric. 
It is, rather, a direct modification of the effective Hamiltonian.

One can say the same for the inclusions of spin-orbit terms done in Refs.~\cite{dam:01, dam:08, nag:11}.
Adding the spin-orbit coupling requires a modification of all terms in the metric that are linear in the spin -  or, equivalently, a direct modification of the Hamiltonian through an additional quantity $\Delta H_\text{so}$ (Eq.~(4.16) of Ref.~\cite{dam:08}).

\item[ii)] Changing the Hamiltonian itself rather than the metric is of course not unreasonable.
Since the motion of a spinning particle is non-geodesic, there is no reason to believe that the dynamics of spinning bodies can be accurately described by geodesics in an effective metric.
One should in principle not worry about intervening on the structure of the effective Hamiltonian itself.
\end{itemize}

In view of these remarks, and noting, in addition, that the effective Hamiltonian depends on the spin squared only through the scalars $\cz^2$, $\ncz^2$, $\pcz^2$ and $\ncz \pcz$, we see a natural way to generalize (\ref{eq:al_prescr}).
We treat the spin differently whether it is contracted with itself, $\bm{n}$ or $\p$, and propose the following type of replacements in Eq.~(\ref{eq:heffexpl}):

\begin{widetext}

\begin{subequations}
\label{eq:prescr}
	\begin{alignat}{4}
	\label{eq:chiprescr}	\cz^2 	 	&\to 	& \, \cesq	& =&&\,	\cz^2 		 + \frac{1}{c^2}\bigg[ z_{11}^{(\chi)} \cu^2 + z_{22}^{(\chi)} \cd^2  +z_{12}^{(\chi) }\cu \cdot \cd \bigg] \\
	\label{eq:nprescr}	\ncz^2	 	&\to 	& \, \ncesq	& =&&\,	\ncz^2		 + \frac{1}{c^2}\Bigg[ z_{11}^{(n)}\ncu^2  +z_{22}^{(n)}\ncd^2 +z_{12}^{(n)}\ncu \ncd \Bigg] \\
	\label{eq:pprescr}	\pcz^2	 	&\to 	& \, \pcesq	& =&&\,	\pcz^2		 + \frac{1}{c^2}\Bigg[ z_{11}^{(p)}\pcu^2  +z_{22}^{(p)}\pcd^2 + z_{12}^{(p)} \pcu \pcd \Bigg] \\
	\label{eq:npprescr}	\ncz\pcz 	&\to 	& \, \npcesq	& =&&\,	 \ncz\pcz	 + \frac{1}{c^2}\Bigg[ z_{11}^{(np)} \ncu\pcu  + z_{22}^{(np)} \ncd\pcd  \nonumber \\
						&	& 	 	&  &&					        + \frac{1}{2} \Big( z_{12}^{(np)} \ncu \pcd  + z_{21}^{(np)}  \pcu \ncd \Big) \Bigg],
	\end{alignat}
\end{subequations}
\end{widetext}
where

\begin{equation}
	z_{ab}^{(\text{x})} \equiv \left( a_{ab}^{(\text{x})}\p^2 + b_{ab}^{(\text{x})}\np^2 + \frac{c_{ab}^{(\text{x})}}{r}  \right),
\end{equation}
the symbol $(\text{x})$ corresponding to $(\chi)$, $(n)$, $(p)$ or $(np)$. 
Recall that the functions $\hat{\Delta}_r$ and $\hat{\Delta}_t$ depend on $\cz^2$, and thus need to be transformed according to (\ref{eq:chiprescr}).
The effective Hamiltonian that results applying (\ref{eq:prescr}) onto $\hhef^\text{old}$ will be simply denoted as $\hhef$.
 
The coefficients $a_{ab}^{(\chi)}$, $b_{ab}^{(\chi)}$ and $c_{ab}^{(\chi)}$ are already known, their explicit expression being given by Eq.~(5.12) of Ref.~\cite{bal:13} (where the label $(\chi)$ had not been used).
In particular, since the $b_{ab}^{(\chi)}$'s vanish, (\ref{eq:chiprescr}) is consistent with Eq.~(\ref{eq:dchieff}).

Determining the $z_{ab}^{(\text{x})}$'s from the Hamiltonian in ADM coordinates cannot be done without finding, simultaneously, the generating function $\hat{G}_\text{ss}^\text{NLO}$ of the corresponding canonical transformation (see Eq.~(\ref{eq:3dtransf})).
We are looking for a sufficiently general ansatz that implements its already known test-mass limit (\ref{eq:GssNLOKerr}), and that mantains, in addition, the symmetry under exchange of the labels 1 and 2.
The searched canonical tranformation may have the following form:

\begin{widetext}
	\begin{alignat}{1}
	\label{eq:GssNLO}
		\hat{G}_\text{ss}^\text{NLO}=\frac{1}{c^6 r} \Bigg\{ 	&  \quad \np \bigg[ \left(\zeta_{11}^{(\chi)}- \frac{1}{2r}\right) \cu^2	+ \left(\zeta_{11}^{(n)}- \frac{1}{2r} \right) \ncu^2 + \delta_{11}^{(p)} \pcu^2 \bigg] 	+ \left( \zeta_{11}^{(np)} + \frac{1}{r} \right) \ncu \pcu  \nonumber \\
									& +\np \bigg[ \left(\zeta_{22}^{(\chi)}- \frac{1}{2r}\right) \cd^2	+ \left(\zeta_{22}^{(n)}- \frac{1}{2r} \right) \ncd^2 + \delta_{22}^{(p)} \pcd^2 \bigg] 	+ \left( \zeta_{22}^{(np)} + \frac{1}{r} \right) \ncd \pcd  \nonumber \\
									& + \np \Big[ \zeta_{12}^{(\chi)} (\cu \cdot \cd) + \zeta_{12}^{(n)} \ncu \ncd    + \delta_{12}^{(p)}\pcu \pcd \Big] \nonumber \\
										& + \frac{1}{2} \Big( \zeta_{12}^{(np)}\ncu \pcd + \zeta_{21}^{(np)}\pcu \ncd \Big) \Bigg\}, 
	\end{alignat}
\end{widetext}
where

\begin{equation}
 \zeta_{ab}^{(\text{x})}  \equiv \left( \alpha_{ab}^{(\text{x})}\p^2 + \beta_{ab}^{(\text{x})}\np^2 + \frac{\gamma_{ab}^{(\text{x})}}{r}  \right)
\end{equation}
for $(\text{x}) = (\chi),(n)$ and $(np)$.
The $\delta_{ab}^{(p)}$'s are, instead, constant coefficients.
Notice that a canonical transformation of this type applies an infinitesimal rotation on the spins according to 
\[ \bm{\chi}_a' = \bm{\chi}_a + \left( \frac{\partial G }{\partial \bm{\chi}_a} \times \bm{\chi}_a \right). \] 

In our specific case, the spins are left invariant under the constraint of aligned spins and equatorial orbits:
\[ \left( \frac{\partial G_\text{ss}^\text{NLO}}{\partial \bm{\chi}_a} \times \bm{\chi}_a \right)\bigg |_\textrm{al}  = 0.\]
Thus, nonprecessing orbits are preserved under the transformation given by (\ref{eq:GssNLO}), which is a consistency requirement for the approach we are following.

\subsection{The general solution}
In order to determine all coefficients, we first explicitly calculate the transformations given by Eqs.~(\ref{eq:1sttransf})-(\ref{eq:3dtransf}).
All needed expressions are already collected in Ref.~\cite{bal:13}, and specifically: $\hat{H}_\text{o}^\text{N(ADM)}$ can be found in Eq.~(2.3) there, $\hat{H}_\text{o}^\text{1PN(ADM)}$ in Eq.~(2.4); $\hat{H}_\text{ss}^\text{LO(ADM)}$ in Eq.~(2.7), $\hat{H}_\text{ss}^\text{NLO(ADM)}$ in Eq.~(2.9);  $\hat{G}_\text{o}^\text{1PN}$ in Eq.~(3.3), and $\hat{G}_\text{ss}^\text{LO}$ in Eq.~(3.5).
We need to rearrange two of these formulae, namely $\hat{H}_\text{ss}^\text{NLO(ADM)}$ and $\hat{G}_\text{ss}^\text{LO}$, that had been originally written in a form that is not suitable for our purpose. 
In Eq.~(2.9b) of Ref.~\cite{bal:13}, it appears the  scalar $\left(\left( \p \times \cu \right) \cdot \bm{n}\right) \left(\left( \p \times \cd \right) \cdot \bm{n}\right)$.
Using the identity (\ref{eq:mixed_pr_id}) (but with the vector $\cu + \cd $ instead of $\cz$) it is easy to show that it can be decomposed as

\begin{widetext}
\begin{align}
	\big(\left( \p \times \cu \right) \cdot \bm{n}\big) \big(\left( \p \times \cd \right) \cdot \bm{n}\big)= 
		& \big(\p^2 - \np^2\big)(\cu \cdot \cd) - \p^2 \ncu \ncd - \pcu \pcd \nonumber \\
		& + \np \big( \pcu \ncd + \ncu \pcd \big).
\end{align}
Eq.~(2.9b) of Ref.~\cite{bal:13} then becomes
\begin{align}
\hat{H}_{\text{S}_1\text{S}_2}^\text{NLO} = 
	& \frac{3 \nu}{r^3}\bigg[ -\left(\frac{1}{2}+ \frac{\nu}{3} \right) \bm{p}^2 (\bm{\chi}_1 \cdot  \bm{\chi}_2)
				+\left(1-\frac{\nu}{4}\right)(\bm{n}\cdot \bm{p})^2 (\bm{\chi}_1 \cdot  \bm{\chi}_2) \nonumber \\
	& +\left(\frac{1}{2}+ \frac{3}{4}\nu \right)\bm{p}^2 (\bm{n} \cdot \bm{\chi}_1) (\bm{n} \cdot \bm{\chi}_2) 
		+\frac{5}{2}\nu  (\bm{n}\cdot \bm{p})^2 (\bm{n} \cdot \bm{\chi}_1) (\bm{n} \cdot \bm{\chi}_2) \nonumber \\ 
	& +\left(\frac{1}{2} + \frac{\nu}{6} \right)(\bm{p} \cdot \bm{\chi}_1) (\bm{p} \cdot \bm{\chi}_2) 
		-\left(1 + \frac{\nu}{4} - \frac{\nu}{2} \frac{m_1}{m_2} \right)(\bm{n}\cdot \bm{p}) (\bm{n} \cdot \bm{\chi}_1) (\bm{p} \cdot \bm{\chi}_2)\nonumber \\ 
	& -\left(1 + \frac{\nu}{4}-\frac{\nu}{2}\frac{m_2}{m_1} \right)(\bm{n}\cdot \bm{p}) (\bm{p} \cdot \bm{\chi}_1) (\bm{n} \cdot \bm{\chi}_2)\bigg] 
		+ \frac{\nu}{r^4} \bigg[ 6  (\bm{\chi}_1 \cdot  \bm{\chi}_2)-12  (\bm{n} \cdot \bm{\chi}_1) (\bm{n} \cdot \bm{\chi}_2) \bigg].
\end{align}
\end{widetext}
Furthermore, using basic vector identities, Eq.~(3.5) of Ref.~\cite{bal:13} is simplifed as follows:
\newline

\begin{alignat}{2}
	\label{eq:2_gen_func}
		\hat{G}_{\text{ss}}^\text{LO}
			&= &&- \frac{1}{c^4}\frac{1}{2\,r^2}\Big \{ \big [ \bm{\chi}_0^2- (\bm{\chi}_0 \cdot \bm{n})^2 \big ] (\bm{r} \cdot \bm{p})\nonumber\\
				& &&+  \big(\bm{\chi}_0 \cdot \bm{n}\big)\left(\bm{r} \times \bm{p}\right)\cdot \big(\bm{\chi}_0 \times \bm{n}\big) \Big \} \nonumber \\
			&= &&-\frac{1}{c^4}\frac{1}{2\,r}\Big[ \np \cz^2 - \ncz \pcz \Big].
\end{alignat}

After all transformations (\ref{eq:1sttransf})-(\ref{eq:3dtransf}), the resulting $H_\text{ss}^{\text{NLO}\prime \prime \prime}$ must be equated to the corresponding term $H_\text{ss}^\text{NLO(EOB)}$ obtained by a PN  expansion of the EOB Hamiltonian 

\begin{equation}
	\hat{H}_\text{EOB} = \frac{1}{\nu}\sqrt{ 1 + 2 \nu \left( \frac{\hhef}{\mu} -1 \right)}.
\end{equation}
The expansion can be done simply replacing $r \to \overline{r} / \varepsilon^2$, $\p \to \varepsilon \, \overline{\p}$ and performing a Taylor series in the small number $\varepsilon$.
$H_\text{ss}^\text{NLO(EOB)}$ is then defined as the part proportional to  $\varepsilon^8$ which is quadratic in the spins.
Finding a solution for the equation 
\begin{equation}
H_\text{ss}^{\text{NLO} \prime \prime \prime}\left( \bm{r}, \p \right) = H_\text{ss}^\text{NLO(EOB)}\left( \bm{r}, \p \right)
\end{equation}
is equivalent to solving an inhomogeneous system of 57 linear equations (18 for the spin(1)-spin(1) combination, 18 for the spin(2)-spin(2) and 21 for the spin(1)-spin(2) one), with 72 variables.
This means that, if the system admits a solution, there will be at least 15 undetermined variables, that, as we shall see, will play the role of gauge coefficients.
Notice that, because of the symmetry under exchange of the particle label 1 and 2, the system can be reduced to 39 equations and 50 variables. 
The general set of equations is solved by:

\begingroup
\allowdisplaybreaks

\begin{align}
 \alpha_{11}^{(\chi)}	&=	 \frac{11 \nu ^2}{32}+\frac{3 \nu ^2}{4} \frac{m_1}{m_2} \nonumber \\
 \beta_{11}^{(\chi)}	&=	 0 \nonumber\\
 \gamma_{11}^{(\chi)}	&=	 \frac{5 \nu}{4} + \left(\frac{\nu}{2} - \frac{\nu^2}{4} \right) \frac{m_2}{m_1} \nonumber\\
\alpha_{11}^{(n)}	&= 	0 \nonumber\\
\beta_{11}^{(n)}	&= 	0 \nonumber\\
 \alpha_{12}^{(\chi)}	&=	 \frac{11 \nu ^2}{16}+\frac{\nu }{2} \nonumber\\
  \beta_{12}^{(\chi)}	&= 	0 \nonumber\\
 \gamma_{12}^{(\chi)}	&= 	-\frac{\nu ^2}{2} \nonumber\\
\alpha_{12}^{(n)}	&=	 0 \nonumber\\
 \beta_{12}^{(n)}	&= 	0. 
\end{align}

\begin{widetext}
\begin{align}
\label{eq:gen_sol}
  a_{11}^{(\chi)} 	&=	 -\frac{11 \nu ^2}{16}-\frac{3 \nu^2}{2} \frac{m_1}{m_2} \nonumber\\
 b_{11}^{(\chi)}	&=	 0 \nonumber\\
 c_{11}^{(\chi)}	&=	 -\frac{29 \nu ^2}{16}-\frac{3 \nu^2}{2} \frac{m_1}{m_2} \nonumber\\
 a_{12}^{(\chi)}	&=	 -\nu -\frac{11 \nu ^2}{8} \nonumber\\
 b_{12}^{(\chi)}	&=	 0 \nonumber\\
 c_{12}^{(\chi)}	&=	 -\nu +\frac{19 \nu ^2}{8} \nonumber\\
 a_{11}^{(n)}		&=	 -\frac{19 \nu }{4}+\frac{39 \nu ^2}{8}+ \left(-\frac{\nu }{2}+\frac{15 \nu ^2}{4}\right) \mdu +\gamma_{11}^{(n)}-\alpha_{11}^{(np)} \nonumber\\
 b_{11}^{(n)}		&=	 \frac{5 \nu }{4}+\frac{15 \nu ^2}{2}+ \left(\frac{5 \nu }{2}+\frac{15 \nu ^2}{4}\right) \mdu -5 \gamma_{11}^{(n)}- \beta_{11}^{(np)} \nonumber\\
 c_{11}^{(n)}		&=	 -\frac{11 \nu }{4}+\nu ^2+ \left(-\frac{\nu }{2}+\frac{7 \nu ^2}{4}\right)\mdu -\gamma_{11}^{(n)}-\gamma_{11}^{(np)} \nonumber\\
 a_{12}^{(n)}		&=	 -2 \nu -\frac{3 \nu ^2}{4}+\gamma_{12}^{(n)}-\frac{1}{2}\left(\alpha_{21}^{(np)}+\alpha_{12}^{(np)}\right) \nonumber\\
 b_{12}^{(n)}		&=	 -5 \nu -\frac{15 \nu ^2}{2}-5 \gamma_{12}^{(n)}- \frac{1}{2} \left(\beta_{21}^{(np)}+ \beta_{12}^{(np)}\right) \nonumber\\
 c_{12}^{(n)}		&=	 -2 \nu +\frac{3 \nu ^2}{2}-\gamma_{12}^{(n)}-\frac{1}{2}\left(\gamma_{21}^{(np)}+\gamma_{12}^{(np)}\right) \nonumber\\
 a_{11}^{(p)}		&=	 2 (\alpha_{11}^{(np)}+\delta_{11}^{(p)}) \nonumber\\
 b_{11}^{(p)}		&=	 2 ( \beta_{11}^{(np)}-2 \delta_{11}^{(p)}) \nonumber\\
 c_{11}^{(p)}		&=	 4 \nu +2 \frac{m_2}{m_1} \nu +\frac{\nu ^2}{2}+2 (\gamma_{11}^{(np)}-\delta_{11}^{(p)}) \nonumber\\
 a_{12}^{(p)}		&=	 \alpha_{21}^{(np)}+\alpha_{12}^{(np)}+ 2\delta_{12}^{(p)} \nonumber\\
 b_{12}^{(p)}		&=	 \beta_{21}^{(np)}+ \beta_{12}^{(np)}- 4 \delta_{12}^{(p)} \nonumber\\
 c_{12}^{(p)}		&=	 -\nu +\nu ^2+\gamma_{21}^{(np)}+\gamma_{12}^{(np)}-2 \delta_{12}^{(p)} \nonumber\\
 a_{11}^{(np)}		&=	 2 (\alpha_{11}^{(np)}- \beta_{11}^{(np)}) \nonumber\\
 b_{11}^{(np)}		&=	 4  \beta_{11}^{(np)} \nonumber\\
 c_{11}^{(np)}		&=	 \frac{13 \nu }{2}+\frac{15 \nu ^2}{4}+ \left(4 \nu +\frac{3 \nu ^2}{2}\right) \mdu +\frac{1}{4} (-8 \gamma_{11}^{(n)}+8 \alpha_{11}^{(np)}+8  \beta_{11}^{(np)}+12 \gamma_{11}^{(np)}+8 \delta_{11}^{(p)}) \nonumber\\
 a_{12}^{(np)}		&=	 2 (\alpha_{12}^{(np)}- \beta_{12}^{(np)}) \nonumber\\
 b_{12}^{(np)}		&=	 4  \beta_{12}^{(np)} \nonumber\\
 c_{12}^{(np)}		&=	 -\frac{5 \nu ^2}{2}-12 \nu ^3 -\left(5 \nu ^2 + 6 \nu ^3\right) \mud -\left(2 \nu ^2+6 \nu ^3\right) \mdu -2 \gamma_{12}^{(n)}+2 \alpha_{12}^{(np)}+2  \beta_{12}^{(np)}+3 \gamma_{12}^{(np)}+2 \delta_{12}^{(p)} \nonumber\\
 a_{21}^{(np)}		&=	 2 (\alpha_{21}^{(np)}- \beta_{21}^{(np)}) \nonumber\\
 b_{21}^{(np)}		&=	 4  \beta_{21}^{(np)} \nonumber\\
	c_{21}^{(np)}		&=	 -\frac{5 \nu ^2}{2}-12 \nu ^3 - \left( 5 \nu ^2+6 \nu ^3\right) \mdu - \left(2 \nu ^2+6 \nu ^3\right)\mud -2 \gamma_{12}^{(n)}+2 \alpha_{21}^{(np)}+2  \beta_{21}^{(np)}+3 \gamma_{21}^{(np)}+2 \delta_{12}^{(p)}.
\end{align}
\end{widetext}
\endgroup

As already mentioned, the coefficients $\alpha_{22}$, $\beta_{22}$, $\gamma_{22}$, $a_{22}$, $b_{22}$ and $c_{22}$ directly follow from the solution above exchanging the particle labels 1 and 2.
In regard to this point, it is worth discussing the Kerr limit $\nu \to 0$. 
In this case, indeed, one has to choose which one of the two spins vanishes - thus somehow breaking the symmetry between them.
For $\nu \to 0$ (and $m_2/m_1 \to 0$) all coefficents $\alpha$, $\beta$ and $\gamma$ must vanish, with the only exception of $\gamma_{22}^{(\chi)}$, $\gamma_{22}^{(n)}$ and $\gamma_{22}^{(np)}$.
The reason of that lies in the form of the canonical transformation (\ref{eq:GssNLO}) and its limit (\ref{eq:GssNLOKerr}).
The price of having enforced the formal symmetry between spin(1) and spin(2), by adding $\nu$-independent terms also to the spin(2)-spin(2) contribution of $\hat{G}_{ss}^\text{NLO}$, is that of generating coefficients $\gamma_{22}$ that do not tend to zero in the Kerr limit (and specifically: $\gamma_{22}^{(\chi)} \to 1/2$, $\gamma_{22}^{(n)} \to 1/2$ and $\gamma_{22}^{(np)} \to -1$, which can be easily verified taking into account that $ \nu \,m_1/m_2 \to 1$).

By contrast, all coefficients $a$, $b$ and $c$ vanish for $\nu \to 0$, as required by consistency. 
In particular, the non-zero limit of $\gamma_{22}^{(\chi)}$, $\gamma_{22}^{(n)}$ and $\gamma_{22}^{(np)}$ is responsible for the convergence towards zero of $a_{22}^{(n)}$, $b_{22}^{(n)}$, $c_{22}^{(n)}$, $c_{22}^{(p)}$ and $c_{22}^{(np)}$, which might not have seemed immediately obvious from Eq.~(\ref{eq:gen_sol}).
Alternatively, to make this convergence more explicit, one could redefine

\begin{align}
	\tilde{\gamma}_{22}^{(\chi)} & \equiv \gamma_{22}^{(\chi)} - \frac{1}{2} = \frac{\nu^2}{2} + \mdu \left( -\frac{\nu}{2} + \frac{\nu^2}{4} \right)\nonumber \\
	\tilde{\gamma}_{22}^{(n)} & \equiv \gamma_{22}^{(n)} - \frac{1}{2} \nonumber\\
	\tilde{\gamma}_{22}^{(np)} & \equiv \gamma_{22}^{(np)} + 1, 
\end{align}
which absorb the $\nu$-independent terms in the spin(2)-spin(2) part of (\ref{eq:GssNLO}), and satisfy $\tilde{\gamma}_{22}^{(\chi)}$, $\tilde{\gamma}_{22}^{(n)}$,$\tilde{\gamma}_{22}^{(np)} \to 0$.
Now we can do the following reformulation:

\begin{alignat}{3}
a_{22}^{(n)}	&= &&\,-\frac{21 \nu ^2}{8}+\left(\frac{\nu }{2}-\frac{15 \nu ^2}{4}\right) \frac{m_2}{m_1} + \tilde{\gamma}_{22}^{(n)}-\alpha_{22}^{(np)}\nonumber\\
 b_{22}^{(n)}	&= &&\,-\left(\frac{5 \nu }{2}+\frac{15 \nu^2}{4}\right) \frac{m_2}{m_1} -5 \tilde{\gamma}_{22}^{(n)}- \beta_{22}^{(np)} \nonumber\\
 c_{22}^{(n)}	&= &&\,-\frac{5 \nu^2}{2}+\left(\frac{\nu }{2}-\frac{7 \nu ^2}{4}\right) \frac{m_2}{m_1} -\tilde{\gamma}_{22}^{(n)}-\tilde{\gamma}_{22}^{(np)}\nonumber\\
 c_{22}^{(p)}	&= &&\,\frac{\nu^2}{2}-2 \nu  \frac{m_2}{m_1}+ 2 \tilde{\gamma}_{22}^{(np)}-2 \delta_{22}^{(p)}\nonumber\\
 c_{22}^{(np)}	&= &&\,\frac{3 \nu^2}{4}-\left(4 \nu +\frac{3 \nu ^2}{2} \right) \frac{m_2}{m_1} -2 \tilde{\gamma}_{22}^{(n)} \nonumber\\
			& &&+2 \alpha_{22}^{(np)}+2\beta_{22}^{(np)}+3 \tilde{\gamma}_{22}^{(np)}+2 \delta_{22}^{(p)}.
\end{alignat}
We stress that, in this case, the formal symmetry with respect to the corresponding $a_{11}$ , $b_{11}$ and $c_{11}$ is not directly visible.

\subsection{Gauge fixing}
The general solution (\ref{eq:gen_sol}) contains 18 gauge parameters, namely $\gamma_{11}^{(n)}$,  $\delta_{11}^{(p)}$, $\alpha_{11}^{(np)}$, $\beta_{11}^{(np)}$, $\gamma_{11}^{(np)}$; $\gamma_{22}^{(n)}$,  $\delta_{22}^{(p)}$, $\alpha_{22}^{(np)}$, $\beta_{22}^{(np)}$, $\gamma_{22}^{(np)}$; and $\gamma_{12}^{(n)}$,  $\delta_{12}^{(p)}$, $\alpha_{12}^{(np)}$, $\alpha_{21}^{(np)}$, $\beta_{12}^{(np)}$, $\beta_{21}^{(np)}$, $\gamma_{12}^{(np)}$, $\gamma_{21}^{(np)}$.
An appropriate choice of them can be used to simplify the 30 coefficients $a_{ab}^{(\text{x})}$, $b_{ab}^{(\text{x})}$ and $c_{ab}^{(\text{x})}$ (for $(\text{x})= (n),(p),(np)$), that are not uniquely determined. 
One could try to impose the so-called Damour-Jaranowski-Sch\"afer (DJS) gauge (see e.g. Refs.~\cite{dam:08,nag:11}) which would consist in eliminating those new terms proportional to $\bm{p}^2$, i.e., all coefficients $a_{ab}^{(\text{x})}$.
We cannot, however, make all of them vanish: there is no way, within the method we have followed, to impose the DJS gauge.
Notice that this does not mean that the DJS gauge is a bad choice in general: in the spin-orbit sector, it is a very useful gauge, independently of how the spin-spin sector looks like.  Moreover, the impossibility of imposing it in  this paper is strictly related with the type of prescription we have followed.
It would be interesting to investigate, in a future work, if a different method for including NLO spin-spin terms would allow to impose a DJS gauge in the spin-spin sector too.

Here, we nevertheless wish to give an example of gauge fixing, and choose an alternative approach.
We can, for example, search a gauge for which the maximal number of coefficients can be set to zero.
One finds that at most 24 of them can vanish, the remarkable fact being that this happens for a unique gauge choice, and namely when the non-zero coefficients are $a_{11}^{(n)}$, $c_{11}^{(n)}$, $a_{22}^{(n)}$, $c_{22}^{(n)}$, $a_{12}^{(n)}$ and $c_{12}^{(n)}$.
In this case, the gauge coefficients are fixed as follows:
\begingroup
\allowdisplaybreaks
\begin{align}
 \gamma_{11}^{(n)} & =  \frac{\nu}{4} + \frac{3}{2}\nu^2 + \left(\frac{\nu}{2} + \frac{3}{4}\nu^2 \right)\frac{m_2}{m_1} \nonumber\\
 \alpha_{11}^{(np)}& = 0 \nonumber\\
  \beta_{11}^{(np)}& = 0 \nonumber\\
 \gamma_{11}^{(np)}& = -2\nu - \frac{\nu^2}{4} - \nu \frac{m_2}{m_1} \nonumber\\
 \delta_{11}^{(p)} & = 0 \nonumber\\
 \gamma_{12}^{(n)} & = -\nu -\frac{3 \nu ^2}{2} \nonumber\\
 \alpha_{21}^{(np)}& = 0 \nonumber\\
  \beta_{21}^{(np)}& = 0 \nonumber\\
 \gamma_{21}^{(np)}& = \frac{\nu ^2}{2}+\nu ^2\frac{m_2}{m_1}  \nonumber\\
 \alpha_{12}^{(np)}& = 0 \nonumber\\
  \beta_{12}^{(np)}& = 0 \nonumber\\
 \gamma_{12}^{(np)}& = \frac{\nu ^2}{2}+\nu ^2 \frac{m_1}{m_2}  \nonumber\\
 \delta_{12}^{(p)} & = 0.
\end{align}
\endgroup

We may also write
\begin{align}
	\tilde{\gamma}_{22}^{(n)} & = - \left( \frac{\nu}{2} + \frac{3}{4} \nu^2 \right)\mdu \nonumber \\
	\tilde{\gamma}_{22}^{(np)} & = - \frac{\nu^2}{4} + \mdu \nu.
\end{align}
The coefficients of the effective spin squared take the following, remarkably simple form:

\begin{align}
a_{11}^{(\chi)} 	&=	- \left(\frac{11}{16}+\frac{3}{2} \frac{m_1}{m_2}\right)\nu^2  \nonumber\\
c_{11}^{(\chi)}		&=	- \left( \frac{29}{16}+\frac{3}{2}\frac{m_1}{m_2}\right)\nu^2 \nonumber\\
a_{12}^{(\chi)}		&=	-\nu -\frac{11 \nu ^2}{8} \nonumber\\
c_{12}^{(\chi)}		&=	-\nu +\frac{19 \nu ^2}{8} \nonumber\\
a_{11}^{(n)}		&=	-\left(\frac{21}{8} + \frac{9}{2}\frac{m_1}{m_2}\right) \nu^2 \nonumber\\
c_{11}^{(n)}		&=	-  \left(\frac{9}{4}+ \frac{m_1}{m_2}\right)\nu ^2\nonumber\\
a_{12}^{(n)}		&=	-3 \nu -\frac{9 \nu ^2}{4}\nonumber\\
c_{12}^{(n)}		&=	-\frac{3 \nu }{2} + \frac{7 \nu ^2}{2},
\end{align}
which is the main result of this paper.
For clarity purposes, we summarize the whole, new effective Hamiltonian:
\begin{widetext}
\begin{alignat}{2}
	\hhef&= \quad 	&&  \frac{\nu r}{2 \tilde{r}_\text{eff}^4}\left(r^2 + \cesq - \hat{\Delta}_t^\textrm{eff}\right) \bigg[\left( \mud g_S^\text{eff} + g_{S^*}^\text{eff}\right)(\bm{n} \times \p)\cdot \cu+\left( \mdu g_S^\text{eff} + g_{S^*}^\text{eff}\right)(\bm{n} \times \p)\cdot \cd \bigg] \nonumber  \\
	  	& +		&&\left(\frac{\hat{\Delta}_t^\textrm{eff}}{\tilde{r}_\text{eff}^4} \right)^{1/2} \left(r^2 + \ncesq \right)^{1/2} \Bigg[ 1 +  \frac{1}{\left(1 + \frac{\ncesq}{r^2} \right)} \Bigg ( \p^2 + \left(\frac{\hat{\Delta}_r^\textrm{eff}}{r^2}-1\right) \np^2 \nonumber  \\ 
		& - &&  \frac{1}{\tilde{r}_\text{eff}^4}\bigg( 2r^2 - \hat{\Delta}_t^\textrm{eff} + \cesq + \ncesq \bigg) \enpcesq \Bigg ) + Q_4(\p) \Bigg]^{1/2},
\label{eq:neweffHam}
\end{alignat}	
\end{widetext}
with
\begin{equation}
\tilde{r}_\text{eff}^4 =\left( r^2 + \cesq \right)^2- \hat{\Delta}_t^\textrm{eff} \left( \cesq - \ncesq \right),
\end{equation}
and with

\begin{alignat}{2}
\enpcesq & =	\: && 2 \np \ncz \pcz - \p^2 \ncesq  \nonumber \\
		& && - \pcz + \left( \p^2 - \np^2 \right) \cesq.
\end{alignat}
The $\Delta$-potentials are

\begin{align*}
\hat{\Delta}_t^\textrm{eff} & = r^2 P^1_3\Big[1 - 2u + 2\nu u^3 + \left(\frac{94}{3} - \frac{41}{32}\pi^2 \right)\nu u^4 + \cesq u^2\Big] \\
\hat{\Delta}_r^\textrm{eff} & = \hat{\Delta}_t^\textrm{eff} \left( 1 + 6\nu u^2 + 2(26-3 \nu)\nu u^3 \right),
\end{align*}
where $P^1_3$ denotes the (1,3)-Pad\'e approximant taken with respect to the variable $u \equiv r^{-1}$.
Notice that, as in Ref.~\cite{bal:13}, we do \emph{not} take the Pad\'e with respect to the variable $r$ contained in $\cesq$.

We recall that the LO effective spin is defined as $\cz = \frac{m_1}{M} \cu + \frac{m_2}{M} \cd$, while the implementation of NLO spin-spin effects reads as follows:
\begin{widetext}
\begin{subequations}
\label{eq:prescrGg}
	\begin{alignat}{4}
	\label{eq:chiprescrGg}	 \cesq	& =	&&\,	\cz^2 		 - \frac{1}{c^2}\bigg\{	&& \bigg[	\left(\frac{11}{16}+\frac{3}{2} \frac{m_1}{m_2}\right)\nu^2 \p^2 + \left( \frac{29}{16}+\frac{3}{2}\frac{m_1}{m_2}\right)\frac{\nu^2}{r}	\bigg]\cu^2 \nonumber\\ 
					&  	&&						&& +\bigg[	\left(\frac{11}{16}+\frac{3}{2} \frac{m_2}{m_1}\right)\nu^2 \p^2 + \left( \frac{29}{16}+\frac{3}{2}\frac{m_2}{m_1}\right)\frac{\nu^2}{r} 	\bigg]\cd^2 \nonumber\\
					& 	&& 		 				&& +\bigg[	 \left(\nu +\frac{11 \nu ^2}{8}\right) \p^2  +\left(\nu -\frac{19 \nu ^2}{8} \right)\frac{1}{r}						\bigg] \cu \cd	\bigg\}
	\end{alignat}
	\begin{alignat}{4}
	\label{eq:nprescrGg}	 \ncesq	& =	&&\,	\ncz^2		 - \frac{1}{c^2}\bigg\{  &&  \bigg[ 	\left(\frac{21}{8} + \frac{9}{2}\frac{m_1}{m_2}\right) \nu^2 \p^2 +  \left(\frac{9}{4}+ \frac{m_1}{m_2}\right)\frac{\nu ^2}{r}			\bigg]\ncu^2 \nonumber\\
					&	&&						&& + \bigg[ 	 \left(\frac{21}{8} + \frac{9}{2}\frac{m_2}{m_1}\right) \nu^2 \p^2 +  \left(\frac{9}{4}+ \frac{m_2}{m_1}\right)\frac{\nu ^2}{r}			\bigg] \ncd^2 \nonumber\\
					&	&&						&& + \bigg[	 \left(3 \nu +\frac{9 \nu ^2}{4}\right)\p^2 + \left(\frac{3 \nu }{2} - \frac{7 \nu ^2}{2}\right)\frac{1}{r} 					\bigg] \ncu \ncd \bigg\}.
	\end{alignat}
\end{subequations}
\end{widetext}
The 25 coefficients that build the rather complex NLO spin-spin Hamiltonian in ADM coordinates get condensed into 12 new contributions in the EOB.
This result is maybe not as striking as in the nonspinning sector (where, at 3PN order, the 11 ADM coefficients are reduced to 3 EOB terms only), still it confirms the notable ability of the EOB in reproducing higher-order effects.


\section{Discussion: ``full'' and ``partial'' inclusion}
\label{sec:discussion}

One of the basic ideas behind the EOB is that of defining a map between real and effective quantities. 
The complicated PN dynamics of spinning bodies, however, has forced a splitting of the concept of ``effective spin'' in the EOB.
Instead of having one single map between $\left( S_1, S_2\right)$ and $S_\text{eff}$, one has to distinguish between linear and squared spin, and even, as shown in this paper, between different contractions of the spin vector with the dynamical variables, and map each component in a different way.
Having thus lost, so to say, the concept of an unique effective spin, one might ask up to which point are we allowed to split the handling of it.
The inclusion of NLO spin-spin effects we have presented in this paper obeys a simple rule: the spin mapping is only diversified when strictly necessary, in other words, the spin terms are as much as possible equally treated. 
We may call this the ``full'' approach, because of its formal and intuitive simplicity.
One could also have followed another line of thought, pursuing the computational efficiency rather than the conceptual unity.
Such an alternative approach would consist in leaving untouched some spin terms of the effective Hamiltonian that actually don't contribute to the NLO spin-spin part of the PN expanded EOB.
For example, the spin-orbit contribution $H_{so}= \Delta H_\text{so} + \beta^i p_i$ of the effective Hamiltonian only generates terms that are odd in the spins, and is thus unrelated to the spin-spin coupling.
We may therefore apply the prescription~(\ref{eq:prescr}) only to the orbital (even in the spins)  part $ H_\text{orb} = \alpha \sqrt{ 1 + \gamma^{ij}p_i p_j}$, leaving $H_\text{so}$ as given by the ``old'' formulation~(\ref{eq:DHbetap}).
We denote this kind of inclusion as ``partial''.
It is clear that this way of proceeding leads to an explicit expression for $H_\text{eff}$ which is shorter and simpler than in the ``full'' case, but it implies, at the same time, an additional and unnecessary fragmentation of the effective spin.
In addition, a problem related to this approach is the fact that the ``partial'' Hamiltonian contains two different Delta potentials, $\Delta_t$ and $\Delta_{t}^\textrm{eff}$.
Since these potentials are defined through a Pad\'e approximant, which may contain poles, the ``partial'' Hamiltonian will in principle show twice as many poles than the ``full'' one, if the spin is nonzero.

Let us finally remark that  the ``full'' and ``partial'' Hamiltonians differ by terms that are odd (and at least cubic) in the spins, and therefore comparing them may help to distinguish the effects intrinsically related to the pure NLO spin-spin coupling from higher-order terms that get automatically resummed into the EOB after the NLO spin-spin inclusion.

\begin{figure}[h!]
	\includegraphics[width= 1.02 \columnwidth]{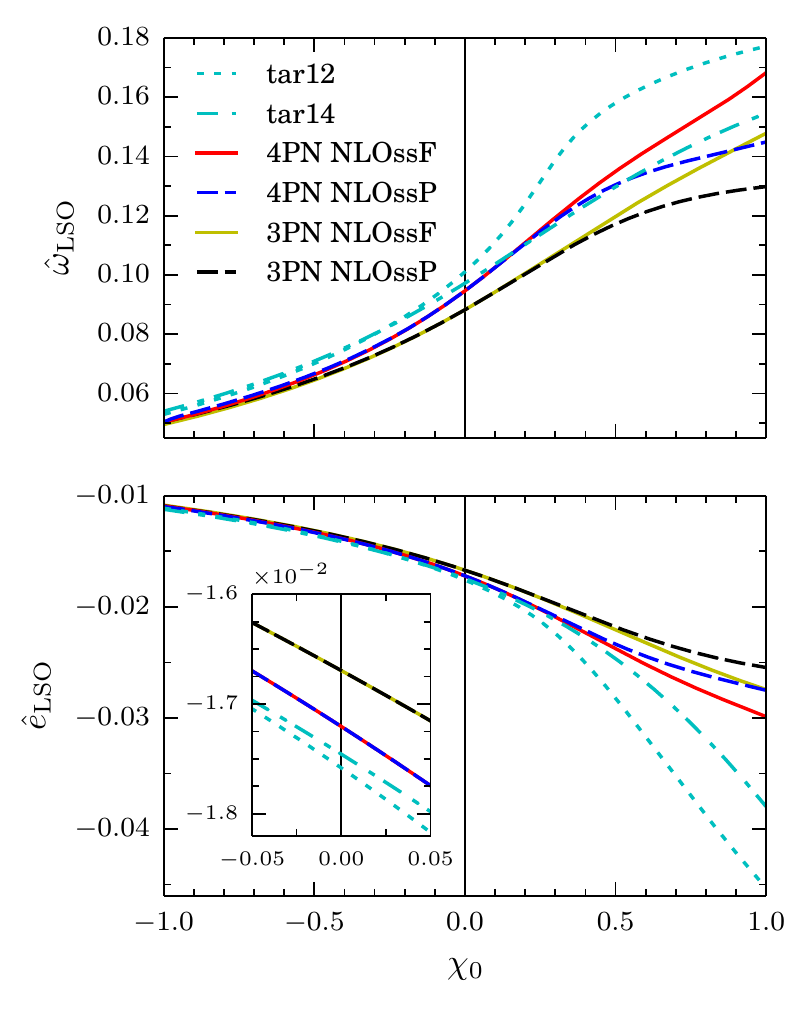}
	\caption{Reduced angular frequency (top) and binding energy (bottom) at the LSO as a function of the effective spin parameter $\chi_0$, in the case of equal masses and equal spins, for different EOB models.} 
	\label{fig:LSO}
\end{figure}

Fig.~\ref{fig:LSO} shows the angular freguency $\hat{\omega}_\textrm{LSO}$ and the binding energy $\hat{e}_\textrm{LSO}$ at the LSO (see also Refs.~\cite{bal:13, bal:13:err}) as a function of the spin parameter $\chi_0$ for different EOB models, in the case of equal masses ($\nu =1/4$).
The curves ``3PN NLOssF'',``3PN NLOssP'', ``4PN NLOssF'' and ``4PN NLOssP'' denote the purely analytical EOB model discussed in this paper and in Ref.~\cite{bal:13}, with NLO spin-spin coupling (``F'' and ``P''indicating the ``full'' or ``partial'' inclusion) and with next-to-next-to-leading order (NNLO) spin-orbit coupling (that is reproduced by the gyro-gravitomagnetic factors calculated in Refs.~\cite{dam:08, nag:11}). 
The label ``3PN'' or ``4PN'' refers to the purely orbital order, resummed with Pad\'e $P^1_3$ or $P^1_4$, respectively.
The 4PN orbital order needs an additional term $\nu \left(a_5^c(\nu) + a_5^{\ln}(\nu) \ln{u} \right)u^5$ (Eq.~(5) of Ref.~\cite{bini:13}) inside of $\hat{\Delta}_t$.  
The figure clearly shows that the effect of the 4PN terms at the LSO is repulsive, since the frequency and the binding energy get increased. 

In Ref.~\cite{bal:13, bal:13:err}, it has already been pointed out that a smaller effective spin squared leads to more bounded orbits.
It is thus not surprising that the ``partial'' inclusion is less bounded than the ``full'' one, as it can be seen in the figure.

For completeness, Fig.~\ref{fig:LSO} also shows two curves generated by a different EOB model.
Here, ``tar12'' and ``tar14'' denote the calibrated model of Refs.~\cite{tar:12,tar:14}.
They both contain analytical information up to LO in the spin squared sector, with a calibration at the NLO level, and to NNLO in the spin-orbit sector, with calibration at next-to-next-to-next-to-leading order (NNNLO).
The main difference between the two models is that ``tar14'' reproduces the 4PN purely orbital order and is calibrated for a varying mass-ratio, while ``tar12'' only contains the 3PN orbital order, and is calibrated for equal masses. 

Since the calibration is done at the waveform level, together with the tuning of some parameters related to the dissipative part, it is difficult to have a precise guess about the real accuracy of the curves ``tar12'' and ``tar14''.
Hoping that the deviation from reality is not so large as to compromise a qualitative discussion of the plot, we can observe that the ``partial'' approach does not seem to show any particular advantage with respect to the ``full'' one.
On the contrary, the ``full'' curves are generally closer to the corresponding calibrated ones.
This might be interpreted as a further argument in favour of the ``full'' approach.
Sistematically replacing, everywhere in the EOB Hamiltonian, terms by their ``effective'' equivalent is thus not only conceptually robust, but seems even to behave well numerically.

A second interesting point is that the difference between our EOB model and the calibrated ``tar'' models  is significantly smaller at the 4PN level than at the 3PN.
In particular, the frequency $\hat{\omega}_\textrm{LSO}$ predicted by ``tar14'' roughly follows the curves ``4PN LSOssF''and ``4PN LSOssP'', and lies in the gap between the two for $\chi_0 \geq 0.5$. 
The maximal deviation between ``4PN'' and ``tar14'' is $6.2\%$ (``P'') and $8.7\%$ (``F''), while between ``3PN'' and ``tar12'' is $24.6\%$ (``F'') and $26.7\%$ (``P''). 
The same is true in the nonspinning regime. 
For $\chi_0=0$, the ``4PN'' value of $\hat{\omega}_\textrm{LSO}$ deviates from ``tar14'' by only $2.6\%$, whereas the difference between ``tar12'' and ``3PN'' is of $12.7\%$. 

In the case of the binding energy, there is a good correspondence for spin equal to zero (the deviations are $1.4\%$ at 4PN and $5.0\%$ at 3PN). 
However, for large spins, the difference is quite large (up to $39.2\%$ (``F'') and $43.5\%$ (``P'') at the 3PN level, and up to $20.9\%$ (``F'')  and $27.1\%$ (``P'') at 4PN). 
Nevertheless, from 3PN to 4PN there is still an improvement up to a factor $\sim 2$.
This fact may strengthen the hope that the need for a calibration becomes less urgent when higher-order analytical terms are included.

\section{Conclusion}

We have proposed a prescription for modifying the EOB Hamiltonian of Ref.~\cite{dam:08}, so that, when expanded in PN terms, it reproduces the correct NLO spin-spin coupling for general precessing orbits.
This is a generalization of the result of Ref.~\cite{bal:13}, where only equatorial orbits with aligned spins had been taken into account. 
The implementation of the correct spin-spin terms is possible after a suitable canonical phase-space transformation of the ADM Hamiltonian.
We have first shown the result in a rather general gauge, with 18 free gauge parameters (that actually reduce to 10 if we require symmetry under exchange of the particles).
Then, a specific gauge is chosen. 
Being impossible, under the type of prescription we have considered, to impose a DJS-type gauge (i.e., to remove all new inclusions of type $\p^2 \chi^2$), we have done the simple choice of looking for a gauge for which the maximal number of the coefficients entering the effective spin squared can be set to zero.
It turned out that there is a unique gauge satisfying this criterion.

In the end, a slightly different approach is taken into account, where the effective spin squared is left at LO in the spin-orbit sector of the effective Hamiltonian. 
This partial inclusion is less repulsive than the full one, and has the unpleasant feature of increasing the number of poles of the Pad\'e approximant by a factor 2. 
A comparison of the (gauge invariant) angular frequency and binding energy, taken at the LSO, with the calibrated EOB models of Refs.~\cite{tar:12,tar:14}, leads to the conclusion that the ``partial'' approach do not show any particular advantage with respect to the ``full'' one. 
By an Occam's razor-like argument, the ``full'' approach, which is conceptually more consistent, may thus be preferable.
As a last thing, the plot also brings the encouraging evidence that, as higher order terms are included, the purely analytical EOB approaches more and more a calibrated model even in the strong field.

\begin{acknowledgments}
We are thankful to Thibault Damour for a very useful discussion.
Moreover, we would also like to thank Andrea Taracchini for having provided us some information.
S.B. is supported by the Swiss National Science Foundation. 
\end{acknowledgments}

\bibliography{biblio}
\bibliographystyle{apsrev}

\end{document}